\documentstyle[12pt]{article}
\begin{document}

\def\beq{\begin{equation}}
\def\eeq{\end{equation}}
\def\bce{\begin{center}}
\def\ece{\end{center}}
\def\bea{\begin{eqnarray}}
\def\eea{\end{eqnarray}}
\def\ben{\begin{enumerate}}
\def\een{\end{enumerate}}
\def\ul{\underline}
\def\ni{\noindent}
\def\nn{\nonumber}
\def\bs{\bigskip}
\def\ms{\medskip}
\def\wt{\widetilde}
\def\brr{\begin{array}}
\def\err{\end{array}}

\hfill DFTUZ / 95/ 02

\hfill hep-th / 9501121

\hfill January, 1995.

\vspace*{15mm}

\begin{center}

{\Large \bf
On the renormalization of  higher derivative \\
two dimensional gravity}

\vspace{30mm}

\renewcommand
\baselinestretch{0.8}
\medskip

{\sc I.L. Shapiro}
\footnote{E-mail address: shapiro@dftuz.unizar.es}
\\
Department of Theoretical Physics, University of Zaragoza, 50009, Spain\\
and\\
Tomsk Pedagogical Institute, 634041 Tomsk, Russia \\

\renewcommand
\baselinestretch{1.4}

\vspace{35mm}

{\bf Abstract}

\end{center}
The fourth derivative models for two dimensional gravity are shown
to be equivalent to the special version of the nonlinear sigma models
coupled to 2d quantum gravity. The reduction consists in the
introduction of the auxiliary scalar fields and can be performed in
an explicit way for both metric and general metric-dilaton cases.
In view of this we can evaluate the structure of possible counterterms
and show that they contains second derivative structures only. The
calculations in the theory with an auxiliary fields require some
special procedure to be applied. We  perform the explicit
calculations in a different gauges and explore the features of the
auxiliary fields.

\vspace{4mm}

\newpage

\ni {\bf 1. Introduction.} \
Recent developments in the field of two dimensional (2d) quantum gravity
(see, for example, \cite{2,3,4,KN,5})
have inspired the interest to the link between
$2d$ and $4d$ gravity theories. In fact, many people regard $2d$ case as the
pattern (or, at least, as  some toy model) for the more realistic four
dimensional quantum gravity. However there is a very essential difference
between these two theories. This difference is constituted, more or less
completely, from two points. First of all in the $2d$ gravity there is no
spin two states (see, for example, \cite{6,npsi} for the discussion of this
in a harmonic gauge, which is most useful in $4d$ gravity) and the metric
can be described by it's conformal factor. Secondly, the loop integrals have
better convergence properties in $2d$ quantum gravity. As a result the
models of $2d$ gravity which are based on the usual action with second
derivative terms only, are renormalizable. In $4d$ the renormalizability
require the fourth derivative terms to be included to the action
\cite{stelle}, that leads to the
loss of unitarity within the standard perturbation scheme (see, for example,
\cite{book} for the introduction to higher derivative  quantum gravity
in $4d$ and references).

Last time there was some interest in the study of the higher derivative
gravity in $2d$  \cite{1984,MO,KNa,ENO,1,NTT,naft}.
The interest to the higher derivative dilaton gravity in $2d$ is
partially caused by the analogy with the sigma model approach to
the massive modes of string, where higher derivatives also appear.
In particular, the general dilaton model with fourth
derivatives has been formulated in Ref.\cite{1}. This model can be regarded
as some tool for investigation of massive string modes.

The study of the one loop renormalization of   the higher derivative
gravity in $2d$ has been started in \cite{ENO,1}
 (see also \cite{NTT,naft}).
The explicit one loop calculation \cite{1}
shows that the divergences which appear in  the general fourth
derivative model have the form of the
second derivative metric - dilaton action.
The same happens in a purely metric model \cite{ENO}. In the last
case the calculations has been performed in a classically equivalent
second derivative model with an auxiliary fields.
Such model has been regarded as
the particular case of the known theory of the dilaton gravity,
and thus the special calculations do not look necessary.

At the same time this way of study leads to some interesting
puzzles. The expression for the one-loop divergences
contains the singular dependence on the auxiliary field and thus it
is not clear what is the link between this result and the starting
$\Lambda+R^2$ model.
On the other hand it turns out that the one-loop calculations
in a harmonic gauge need some special procedure including the
noncovariant terms to be introduced. Such terms exist in the
dimension $d = 2 + \varepsilon$ and the contribution of this terms
disappear when the regularization is removed and the parameter
related with an extra terms tends to zero. The necessity of this
noncovariant procedure leads to conclusion that the higher derivative
$d=2$ theory needs special renormalization procedure including the
addition of some extra noncovariant counterterms \cite{naft}.
All this is rather unnatural since there are no any difficulties
of this sort in conformal gauge.

The goal of this letter is to solve the above puzzles.
It is shown, how the covariant calculations can be performed without
extra noncovariant terms. Next, we demonstrate
that  quantum
theory with an auxiliary fields has to be modified as compared
with the ordinary dilaton gravity. \footnote{The theories with an
auxiliary fields have been studied in \cite{Coleman}.}
Then,  in the framework of the
modified theory the higher derivative gravity models are shown to be
renormalizable in standard sence in both harmonic and conformal gauges.
The use of conformal gauge enables one to show that only the second
derivative counterterms arise at all loops. This concerns both
$\Lambda+R^2$ model and general dilaton model of \cite{1}.
The last is shown to be
 equivalent to  the $D = 4$ second derivative sigma model
where some of the sigma model coordinates are auxiliary scalars.

The paper is organized as follows. In section 2 we consider the pedagogical
example of higher derivative $2d$ gravity without dilaton. In section 3
we deal with the general model of \cite{1}. The last section is conclusion.
\bs

\ni {\bf 2 Metric model with fourth derivatives.} \
As starting point, let us
consider the higher derivative metric theory.
\beq
S_{g} = \int {d^2}x \sqrt {g} \;\{ \;{1\over m^2} \;R^2+
\Lambda\; \}                                                   \label{1}
\eeq
Here $ m^2,\Lambda $ are dimensional constants. We do not include the
Einstein term because for the $2d$ space without boundaries
it does not affect the renormalization. The action (\ref{1}) may be
supplemented by the action of $N$ scalar fields
\beq
S_m = \int {d^2}x \sqrt {g} \; \frac{1}{2}\; g^{\mu\nu}
\partial_{\mu}\chi^i \partial_{\nu}\chi^i                     \label{1a}
\eeq
where $i = 1...N$. Those fields give contribution to the Einstein
type divergences \cite{24}.
Below we omit the terms, related with $S_m$ in all the
intermediate expressions for the sake of brevity, and take them into account
in the final expressions only.

The above theory has been studied on quantum level in \cite{1984,MO,KNa,ENO,
NTT,naft}. In particular, in the works \cite{ENO,NTT,naft}
the one loop renormalization of the theory has been discussed.
In \cite{ENO,NTT}
the calculation of the one-loop divergences has been performed with the
help of the auxiliary scalar field.
Introducing the auxiliary scalar $\psi$ one can substitute (\ref{1}) by the
second derivative theory with the classical action
\beq
S = \int {d^2}x \sqrt {g} \{ -\frac{m^2}{4}\psi^2 + \psi R +
\Lambda \}                                              \label{2}
\eeq
On classical level the theories (\ref{1}) and (\ref{2}) are equivalent
that can be seen if one take the value of $\psi$ from the corresponding
equation of motion and substitute it back to the action (\ref{2})
or to the dynamical equation for metric.

The theory (\ref{2}) looks like a particular
case of the general action for dilaton gravity
\beq
S_{gen} = \int {d^2}x \sqrt {g} \left\{\; \frac{C_0}{2}
g^{\mu\nu}\;\partial_{\mu}\phi\; \partial_{\nu}\phi + C_1\;\phi R +
V(\phi) \;\right\}                                          \label{2a}
\eeq
where $C_0 = 1,0$ for different conformally
equivalent versions of the theory and $V(\phi)$ is the potential function.
The one loop renormalization in the theory with the action (\ref{2a})
has been investigated in \cite{23,23a,24,25,26}.
In a harmonic type background gauge
$$
S_{gf} = - {C_{1}\over 2} \int {d^2}x \sqrt {g} \; {\chi}_{\mu}
\;{\phi\over{\nu}}\;{\chi}_{\nu}
$$
\beq
\chi_\mu = \nabla_\nu h_\mu^\nu - {1\over2}
\nabla_\mu h - {1\over2}\nu\nabla_\mu\phi - X(\phi)
\nabla_\mu\phi                                        \label{2b}
\eeq
the divergent part of the one - loop effective action has
the form (it does not depend on the choice of $C_0$ as well as on the
gauge parameter $X(\phi)$ )  \cite{23a}
\beq
\Gamma_{div} = \frac{1}{\varepsilon}\;\int {d^2}x \sqrt {g}
\left\{\;A_1(\phi) g^{\mu\nu}\partial_{\mu}\phi \partial_{\nu}\phi
+A_2(\phi)R+ A_3(\phi)    \; \right\}    \label{3}
\eeq
with
\beq
A_1(\phi) = - \frac{1}{\varepsilon}\frac{\nu}{\phi^2}\; \;\;\;\;\;\;\;\;
A_2 = \frac{1}{\varepsilon}\;\frac{24-N}{12} \; \;\;\;\;\;\;\;\;
 A_3(\phi) = \frac{\nu\;V}{C_1\;\phi} + \frac{V'}{C_1}         \label{gen}
\eeq
where $\nu$ is an arbitrary gauge parameter and
$g_{\mu\nu}$ and $\phi$ are background metric and scalar field.
$\varepsilon = (4\pi)^2\;(d-2)$ is the parameter of dimensional
regularization. Note that in the
theory under discussion the propagator has smooth behaviour when
$\varepsilon$ tends to zero, and hence here is no additional effects like
the oversubtraction problem \cite{KN} which may come from
dimensional regularization.
Thus it seems possible to choose the special values of $C_0$ and $V(\psi)$
corresponding to (\ref{2}), substitute them into (\ref{3}), and so obtain the
divergences of the original theory (\ref{1}) \cite{ENO,NTT}
\footnote{In fact in \cite{NTT} the independent calculation has been performed.
Surprisingly the dependence of the gauge parameter $X(\psi)$ has been
found. This dependence contradict to the results of \cite{23a} and of third
reference in \cite{24}. Moreover the detailed analisys shows that such
a dependence leads to the on shall gauge dependence of the one-loop
divergences}.
However on this way one face some difficulties.
The point is that $\psi$ is the auxiliary field and one have to
care about it and especially if the background field method is used. Let us
consider this in details.

Quantum theory is conventially
constructed by means of generating functional in the form of
the path integral. Since $\psi$ is the auxiliary field one
has to avoid the introduction of the external source for this field.
The last can be interpreted as the simple fact that only the diagrams
without external $\psi$ lines are valid. Therefore the path integral
representation of the theory (\ref{2}) is different from the one of
(\ref{2a})
just because some diagramms in the last must be removed. In the theory with
auxiliary field $\psi$ the propagator of this field works on the internal
lines only.
In the framework of the background field method the Green functions
arise as a result of variation of the effective action with respect to
the background fields. Hence if one need to abort the diagramms with
the external $\psi$ lines it is necessary to consider the theory without
background $\psi$ field.
So we see that in the background field method (\ref{1}) corresponds to the
theory with  purely quantum field $\psi$. Let us notice that the same results
from the formal analisys (one can find the details of the background
field method in a numerous papers or in \cite{book}).

{}From the above consideration it follows that it is not possible to obtain
the correct expression for divergences in the theory (\ref{1}) with the
help of (\ref{3}). Since both the gauge fixing term (\ref{2b}) and the
divergences (\ref{3}) contain $\frac{1}{\psi}$ factors they do not have
smooth behaviour in the limit $\psi\rightarrow 0$ and thus this way does not
provide us by correct result. Indeed all the singular terms vanish when
the gauge parameter is choosen in a special way $\nu = 0$. However it is
not clear whether the resulting expression has any link to the original
theory (\ref{1}).

In the rest of this section we calculate the one loop divergences in the
theory (\ref{1}) by three different ways, that is

i)In terms of original
higher derivative theory (\ref{1}) in a harmonic type background gauge,

ii)In terms
of (\ref{2}) in similar gauge,

iii) In terms of (\ref{2}) in conformal gauge.

The last case is especially interesting since it enables one to establish the
structure of  renormalization at higher loops.
\vskip 1mm
i) Calculation in   original higher derivative theory.
According to the background field method we
divide the metric into background
$g_{\mu\nu}$ and quantum $h_{\mu\nu}$ parts as
\beq
g_{\mu\nu}\rightarrow g'_{\mu\nu}= g_{\mu\nu} + h_{\mu\nu},
\;\;\;\;\;\;\;\;\;\;\;
h_{\mu\nu} = {\bar h}_{\mu\nu} + \frac{1}{2}g_{\mu\nu}        \label{3n}
\eeq
and introduce the gauge fixing term of the form
\beq
S_{gf} = - \frac{1}{\alpha}\;
\int {d^2}x \sqrt {g} \;({\nabla}_{\tau} {\bar h}_{\mu}^{\tau}
-\beta {\nabla}_{\mu}h)\;[g^{\mu\nu}\Box +\gamma \nabla^{\mu}\nabla^{\nu}]\;
({\nabla}_{\lambda}{\bar h}_{\nu}^{\lambda} -
\beta {\nabla}_{\nu} h)                                      \label{3b}
\eeq
where $\alpha, \beta, \gamma$ are the gauge fixing parameters.
The one-loop effective action is given by the expression
\beq
\Gamma_{div} = \frac{i}{2}\;Tr\ln {\hat {H}}-i\;Tr\ln{\hat {H}}_{gh}+
\frac{i}{2}\;Tr\ln [g^{\mu\nu}\Box+\gamma\nabla^{\mu}\nabla^{\nu}]  \label{3c}
\eeq
where ${\hat H}$ is the bilinear form of the action $S + S_{gf}$,
${\hat H}_{gh}$ is the bilinear form of the ghost's action and the last term
is the contribution of the weight operator (\ref{2b})
The calculations are  simplified
considerably, if one  choose the values of $\beta$ and $ \gamma$
as $\beta = \frac{\alpha}{2m^2}\left( \frac{\alpha}{m^2} - 1 \right)$
and $\gamma = - \frac{\alpha}{m^2}$. Then, after some algebra, we arrive at
the following bilinear part of $S + S_{gf}$
$$
S^{(2)} + S_{gf} = \int {d^2}x \sqrt {g} \;\left\{
{\bar h}^{\mu\nu}\;\left[- \frac{2\beta\gamma}{\alpha} - \frac{1}{2m^2}
\right]R\;{\nabla}_{\mu}{\nabla}_{\nu}\;h+
\right.
$$
$$
\left.
{\bar h}^{\mu\nu}
\left[\frac{1}{\alpha}\;g_{\nu\beta}{\nabla}_{\mu}{\nabla}_{\alpha}\Box
-\left(\frac{1}{2m^2} +\frac{1}{2\alpha}\right)R\;
g_{\nu\beta}{\nabla}_{\mu}{\nabla}_{\alpha} -
\frac{1}{2m^2}\;R^2\;\delta_{\mu\nu ,\alpha\beta}
-\frac{m^2}{4}\;\Lambda\;\delta_{\mu\nu ,\alpha\beta}\right]
{\bar h}^{\alpha\beta}+
\right.
$$
\beq
\left.
h\;\left[\left(
\frac{\beta^2\; (m^2 - \alpha)}{\alpha m^2} + \frac{1}{4} \right)
\Box^2 +
\left( \frac{\beta^2}{2\alpha} - \frac{1}{4m^2} \right)R
\right]\;h
\right\}                                                   \label{3d}
\eeq
Let us now use the two - dimensional identity
\beq
{\bar h}^{\mu\nu}\;X\;
\;\left[ \;g_{\nu\beta}{\nabla}_{\mu}{\nabla}_{\alpha}\right]
{\bar h}^{\alpha\beta} =
\frac{1}{2} \; {\bar h}^{\mu\nu}\;X\;
\;\left[  \delta_{\mu\nu ,\alpha\beta}\Box -   \delta_{\mu\nu ,\alpha\beta}R
\right] {\bar h}^{\alpha\beta}                            \label{3e}
\eeq
advocated in \cite{6,npsi}. Here $X$ is an arbitrary scalar quantity.
Since the prove of (\ref{3e}) use only the
tracelessness of ${\bar h}^{\alpha\beta}$ it is not affected by the action of
$X = \Box$. Then, after simple rescaling of the fields
${\bar h}^{\alpha\beta}, \; h$, the bilinear form ${\hat {H}}$
has the form of the minimal higher derivative operator
\beq
{\hat {H'}} = {\hat {1}}\; \Box^2 +{\hat {L}}^{\alpha\beta\nu}
\nabla_\alpha\nabla_\beta\nabla_\nu
+{\hat {V}}^{\alpha\beta}\nabla_\alpha\nabla_\beta
+{\hat {N}}_{\alpha}\nabla_\alpha + {\hat {U}}         \label{3f}
\eeq
with ${\hat {L}}^{\alpha\beta\nu} = 0$. The divergences of
$Tr\ln {\hat {H'}}$ has been derived in \cite{23a}.
\beq
Tr\ln {\hat {H}}\|_{div}= - \frac{2}{\varepsilon}\;tr
\left(\;\frac{1}{2}\;{\hat {V}}_{\alpha}^{\alpha} -
\frac{1}{16}\;{\hat {L}}^{\alpha\beta\nu}\;{\hat {L}}_{\alpha\beta\nu}
-\frac{3}{32}{\hat {L}}^{\alpha}\;{\hat {L}}_{\alpha}
\;\right)                                              \label{3g}
\eeq
If one use (\ref{3g}) and take into account the contributions of ghosts
and of the weight operator, the dependence on the gauge parameter $\alpha$
is droped out and the divergences of the theory (\ref{1}) are found to be
\beq
\Gamma_{div} = \frac{1}{\varepsilon}\int {d^2} x \sqrt {g}
\;\left( \;\frac{24-N}{12}\; \right)\;R                    \label{4}
\eeq
Thus the only one
counterterm which appears in higher derivative theory (\ref{1})
is (\ref{4}).

Some note is in order. From the above consideration we learn that in two
dimensional gravity any second or fourth order differential operator in
${\bar h}{\bar h}$ sector is minimal due to identity (\ref{3e}). As a
consequence the calculation in harmonic type gauge in a higher
derivative $d=2$ gravity is possible without
introduction of the extra noncovariant term like
$\xi {\bar {h}}^{\mu\nu}\Box(R_{\mu\nu} -\frac{1}{2}Rg_{\mu\nu})$
that has been done in \cite{1,naft}. From what follows that both the
loop calculations and renormalization procedure can be performed in
two dimensions and there is no any need to introduce an extra noncovariant
structures in $d = 2+\varepsilon$ \cite{naft}.

\vskip 1mm

ii) Now we shall look how all this works in the classically equivalent
theory (\ref{2}) when the auxiliary field $\psi$ is introduced. Since
the auxiliary
field is purely quantum we must introduce the background metric only,
as a result the calculations are tiny.
Then,  performing the splitting of the metric
 as in (\ref{3n}) we choose the gauge fixing term in
the form
\beq
S_{gf} = \frac{1}{\alpha}\;\int {d^2}x \sqrt {g}
\;({\nabla}_{\tau} {\bar h}_{\mu}^{\tau}
+\frac{\alpha}{2} {\nabla}_{\mu} h)
({\nabla}_{\lambda} {\bar h}^{\mu\lambda}
+\frac{\alpha}{2} {\nabla}^{\mu} h)                              \label{3h}
\eeq
Then, after use of (\ref{3e})
we find that the bilinear (with respect to quantum fields $h_{\mu\nu}, \psi$ )
part of the expression $S + S_{gf}$ has the form
\beq
S^{(2)} + S_{gf} = - \int {d^2}x \sqrt {g} \;\left\{ \frac{1}{2\alpha}
{\bar h}^{\mu\nu}\;(\Box - R + \frac{\alpha}{2}\;\Lambda ){\bar h}_{\mu\nu}
 + \frac{1}{2}\;\psi\Box h + \frac{\alpha}{4}\;\psi\Box\psi
 + \frac{m^2}{4}\;\psi^2
\right\}                                                   \label{3i}
\eeq
The above expression enables us to use standard algorithm for the
minimal second order operators. Taking into account the contribuion of
gauge ghosts we finally obtain the
divergences of the theory (\ref{2}) with an auxiliary field $\psi$
\beq
\Gamma_{div} = \frac{1}{\varepsilon}\int {d^2} x \sqrt {g}
\;\left(\;\frac{24-N}{12}\; R-\frac{\alpha}{2}\Lambda\right) \label{3j}
\eeq
The last expression differs from (\ref{4}) by the cosmological term.
The source of this deviation is the different gauge fixing conditions.
In fact (\ref{3j}) are the  divergences of the theory (\ref{2}) with
zero background $\psi$. For zero $\psi$ the equation of motion for the
theory (\ref{2}) reads $\Lambda = 0$, and two expressions
coinside on this "mass shell". The problem is that $\Lambda = 0$ does
not correspond to equations of motion of the original theory (\ref{1}).
 Thus the defect of the
divergences (\ref{3j}) is caused by the calculational scheme, and we conclude
that in the harmonic gauge two theories (\ref{1}) and (\ref{2}) are not
completely equivalent on quantum level even if we treat auxiliary fields
in a correct way.
\vskip 1mm
iii) The story is much more simple if we apply the conformal gauge.
Let us write
\beq
g_{\mu\nu}={\bar {g}}_{\mu\nu}\;e^{2\sigma}         \label{5}
\eeq
where $\sigma$ is quantum field  and ${\bar {g_{\mu\nu}}}$
is the background metric. The scalar curvature is transformed as
\beq
R=e^{-2\sigma}[{\bar {R}} - 2{\bar {\Box}} \sigma]              \label{6}
\eeq
and thus (\ref{2}) is reduced to the ordinary sigma-model action
(see also the paper of Russo and Tseytlin \cite{24} where similar
representation has been used in second derivative $2d$ dilaton gravity.)
\beq
S = \int {d^2}x \sqrt{{\bar g}} \{\; {\bar g}^{\mu\nu} G_{ab}\;
\partial_{\mu}X^a \;\partial_{\nu}X^b +B(X){\bar R}+T(X)  \;\}    \label{7}
\eeq
where
\beq
X^{a} = (\psi,\;\sigma),\;\;\;\;\;\;\;
G_{ab} = \left(\matrix{
              0   &1 \cr
              1   &0
\cr}\right),\;\;\;\;\;\;\;
B(X) = \psi + \alpha,\;\;\;\;\;\;\;
T(X) = \Lambda  -\frac{m^2}{4}\;\psi^2\;e^{2\sigma}         \label{8}
\eeq
Thus in conformal gauge the theory (\ref{2}) becomes the
linear sigma model which is known to have the divergences of the
tachyon type, and also the anomalous one of the
Einstein type. This is the divergences structure not only at one,
but also at higher loops. Since in the case of the theory (\ref{1})
only metric is the dynamical field, one can expect for the divergences
of Einstein and cosmological form only. Let us notice that even if we consider
the theory with auxiliary field, and some diagramms have to be omitted, the
superficial index of the remaining diagramms is in an accord with the
power counting of the general theory (\ref{8})
(or (\ref{2a})), and thus our consideration here is correct.

It is important that the linear sigma model calculations can be
performed without mixing of conformal factor $\sigma$ with the auxiliary
field $\psi$ and that the result is not singular when we put $\psi$ equal
to zero.
Since in the case the "tachyon" term is constant for the
zero background $\psi$, such a divergences
are lacking as well. The explicit one-loop calculations show that
the divergences of (\ref{2}) have the form (\ref{4}) just as in original
formulation of the theory (\ref{1}).

\bs

\ni {\bf 3. Metric - dilaton model.} \

Let us now consider a general version of quantum dilaton gravity which
has been recently formulated in \cite{1}. The action of the model has
the form
$$
S= \int d^2x \, \sqrt{g} \, \left\{
a_1 (\varphi)R^2 + a_2 (\varphi)R \Box \varphi+ a_3(\varphi) R
\left( \nabla\varphi \right)^2
+ a_4 (\varphi ) \left( \Box\varphi \right)^2+a_5 (\varphi )
\left( \Box\varphi \right)\left( \nabla\varphi \right)^2+
\right.
$$
\beq
\left.
+a_6 (\varphi)\left( \nabla\varphi \right)^4
+a_7 (\varphi )R+
 a_8 (\varphi ) \left( \nabla\varphi \right)^2
+a_9 (\varphi)
\right\}.                                        \label{9}
\eeq
Here $a_i(\varphi)$ are an arbitrary functions of the dimensionless scalar
field $\varphi$ and $\left( \nabla\varphi \right)^2=
\left( \nabla_\mu \varphi
\right)\left(\nabla^\mu\varphi\right)$.

The one - loop calculations in the model (\ref{9}) have been performed
in \cite{1} for both general case and for $a_2 = a_4 = 0$. Here we
show how the general model (\ref{9}) can be reduced to the second derivative
theory, that enables one to establish the structure of divergences at higher
loops. Moreover we shall derive the one loop divergences
in this second derivative
theory and check the correspondence between two
formulations on quantum level.

Despite the action (\ref{9}) looks rather combersome it is possible to
introduce the auxiliary fields and to reduce
(\ref{9}) to the lower derivative
action. To show this one can rewrite  (\ref{9}) in the following form
$$
S_{dg}=\int d^2x \, \sqrt{g} \, \left\{
a_7 (\varphi )R + a_8 (\varphi ) \left( \nabla\varphi \right)^2
+a_9 (\varphi)+
\right.
$$
\beq
\left.
+a (\varphi) \left[\; R+y(\varphi) \left( \Box\varphi \right)
+z(\varphi)\left( \nabla\varphi \right)^2 \;\right]^2 +
b(\varphi) \left[\;
\left( \Box\varphi \right) + u(\varphi)R+v(\varphi)\left(
\nabla\varphi\right)^2 \;\right]^2
\right\}.                                                \label{10}
\eeq
Here $a (\varphi),  b(\varphi), y(\varphi), z(\varphi), u(\varphi),
v(\varphi)$ are some functions
which are related with $a_{1,...,6}(\varphi)$.
Now we are able to introduce the auxiliary fields $\psi$ and $\xi$
and to rewrite (\ref{10}) in the form
$$
S_{dg}= \int d^2x \, \sqrt{g} \, \left\{
-\frac{1}{4a(\varphi)}\psi^2 -\frac{1}{4b(\varphi)} \xi^2
+\psi \left[ R+y(\varphi) \left( \Box\varphi \right)
+z(\varphi)\left( \nabla\varphi \right)^2  \right]+
\right.
$$
\beq
\left.
+\xi \left[
\left( \Box\varphi \right)
+u(\varphi)R+v(\varphi)\left( \nabla\varphi \right)^2
\right] +a_7 (\varphi )R+
 a_8 (\varphi ) \left( \nabla\varphi \right)^2  + a_9 (\varphi)
\right\}.                                        \label{11}
\eeq
The last action can be regarded as the special case of the nonlinear sigma
model coupled to quantum metric.
The one loop counterterms for the general nonlinear sigma
model coupled to quantum $2d$ gravity have been recently calculated
in \cite{npsi}. Unfortunately
the result of \cite{npsi} can not be applied directly
to the model (\ref{11}) because the auxialary fields $\psi, \xi$ need
the scecial care. Moreover
the metric of the sigma model (without quantum
gravity contributions) is degenerate in the last case. However the method of
 \cite{npsi} is applicable, and thus we can use the statement of equivalence
between covariant and conformal gauges which was proved there at one - loop
level. Therefore it is sufficient to consider the conformal gauge only.
In the conformal gauge the  action (\ref{11}) becomes
 (the argument $\varphi$ and bars are omitted below)
\beq
S_{dg} = \int {d^2}x \sqrt {g} \;\{\; g^{\mu\nu} G_{ab}
\partial_{\mu}X^a \partial_{\nu}X^b +B(X^a)R+T(X^a)\;\}            \label{12}
\eeq
where
$$
X^{a} = (\psi,\;\xi,\;\phi,\;\sigma),\;\;\;\;\;\;\;
G_{ab} = \left(\matrix{
0     &  0     &  - y/2                      &   1    \cr
0     &  0     &  - 1/2                      &   u    \cr
-y/2  & - 1/2  &   (v\xi+z\psi+a_8 -y'\psi)  & u'\xi  \cr
1     &  u     &    u'\xi                    & 0
\cr}\right),\;\;\;\;\;\;\;
$$
\beq
B(X) = \psi + u\xi + a_7,\;\;\;\;\;\;\;
T(X) = \left(\; a_9 -\frac{1}{4a}\;\psi^2 -\frac{1}{4b}\;\xi^2\;\right)
\;e^{2\sigma}                                               \label{13}
\eeq
and the prime stands for the derivative with respect to $\phi$.
According to our analisys in the previous section, on quantum level
the theory (\ref{11}) corresponds to the path integral with
the external sources for the metric  and $\phi$ only, but not to the
auxiliary fields $\psi, \xi$. In the framework of the background field
methos it means that we have both this fields as purely quantum.
However the aborting of the diagrams with an external $\psi$ and $ \xi$
lines does not change the power counting. Therefore the general form
of divergences in the theory (\ref{11}) \footnote{Perhaps with one exception,
that will be  discussed below.} is given by the expression (\ref{gen}).
Indeed the functions $A_{1,2,3}(\phi)$ can depend on
$a (\varphi),  b(\varphi), y(\varphi), z(\varphi), u(\varphi), v(\varphi)$.
Thus the introduction of higher derivative terms is not caused by
the renormalizability, in contrast with the four dimensional gravity.
And so we observe that the result of direct calculations of Ref.\cite{1} is
just that it has to be.

The divergences of the sigma model (\ref{12}) are well known \cite{FT,CFMP}.
However the general result of  \cite{FT,CFMP} is not applicable in our
case since the method of calculation is assumed to preserve the
covariance with respect to the target space reparametrizations and
the corresponding calculational method can mix the auxiliary fields with
$\phi$ and $\sigma$. As a result one can not simply put the background
auxiliary fields equal to zero, and all the consideration is essentially
more complicated.

Since we are interested in the application of the auxiliary fields to the
one - loop calculations, let us consider the example of (\ref{11})
and compare the one loop divergences  with the ones of \cite{1}.
It is reasonable
to apply conformal gauge which has been used in  \cite{1}. Expanding
the conformal factor and $\phi$ as
\beq
\sigma \rightarrow {\bar {\sigma}} + \sigma\;\;\;\;\;\;\;\;\;\;\;\;\;\;
\phi\rightarrow\phi+\tau                                  \label{14}
\eeq
we shall regard $\phi$ and
$g_{\mu\nu}={\bar {g}}_{\mu\nu}\;exp(2{\bar {\sigma}})$
as the background quantities and $\sigma, \tau$ as quantum ones. After
small algebra we obtain the bilinear part of the action (\ref{12})
in the form
\beq
S^{(0)} = \int {d^2}x \sqrt {g} \;\{\; G_{ab}
\nabla_{\mu} X^a \nabla^{\mu} X^b + A_{ab} (\nabla^{\mu}\phi)X^a
\nabla_{\mu} X^b + B_{ab}\;\}                              \label{15i}
\eeq
where $ G_{ab}$ is given by (\ref{13}) with $\xi = \psi = 0$ and
$$
A_{ab}= \left(\matrix{
0          &  0              &     0   &   0  \cr
2a''_7     &  2a_8           &  -  y'  &   0 \cr
0          & 2z - 2y'        &     0   &   0  \cr
2u'        &  2v             &     0   &   0
\cr}\right),\;\;\;\;\;\;\;
$$
\beq
B_{ab}= \left(\matrix{
2a_9          &  0                             &     0   &   0  \cr
2a'_9 & 1/2(a''_9+a''_7R+a''_8(\nabla\phi)^2)  &  (z-y')(\nabla\phi)^2)&
 u'R+v'(\nabla\phi)^2 \cr
0          & 0        &   -1/4a   &   0  \cr
2u'        &  2v             &     0   &  -1/4b
\cr}\right)                                              \label{16i}
\eeq
The above expression leads to the standard bilinear form of the action. The
one loop divergences can be easily estimated within the Schwinger-DeWitt
method and finally are given by expression
$$
\Gamma_{div} = \frac{1}{\varepsilon}\;\int {d^2}x \sqrt {g}
\;tr \left\{ G^{-1}M - \frac{1}{4} G^{-1}\;LG^{-1}\;L + \frac{1}{6}\;R
\; \right\}
$$
\beq
L(\nabla_\mu\phi) = (\nabla_\mu G) + \frac{1}{2}(A^T - A)(\nabla_\mu\phi)
\;\;\;\;\;\;\;\;\;
M = - \frac{1}{2}(B+B^T)+ \frac{1}{2}A^T(\Box\phi)+ \frac{1}{2}(\nabla_\mu A^T)
(\nabla_\mu\phi)                                           \label{17}
\eeq
Substituting (\ref{16i}) into (\ref{17}) we obtain
$$
\Gamma_{div} = \frac{1}{\varepsilon}\;\int {d^2}x \sqrt {g}
\;\left\{
\frac{2u'}{1-uy}\;R -
\frac{u(a'_7 + ua_8)}{a(1-uy)^2} - \frac{a_8+ya'_7}{(1-uy)^2}
+\frac{2uz-3uy'-u'y-2v}{1-uy}\;(\Box\phi)+
\right.
$$
$$
\left.
\frac{1}{(1-uy)^2}(\nabla\phi)^2\;[(uy''+yu'')(1-uy)+
\right.
$$
\beq
\left.
+u^2(y'')^2+y^2(u'')^2 + 2(z-y')(zu^2 - u')+2v(u'y+uy'-2uv+z)
\; \right\}                                                 \label{18}
\eeq
Now, substituting the relations between $a_i(\phi)$ (\ref{9}) and
$a(\phi), b(\phi), y(\phi), z(\phi), u(\phi), v(\phi)$ (\ref{10})
into the corresponding expression from \cite{1} we find that both results
are in a good relation with each other. One can regard this as an
additional check of our treating of the auxiliary fields.
Note that applying the standard result \cite{FT,CFMP}
for the sigma model (\ref{12})
we arrive at different expression. This can be easily seen from the
"tachyon" sector already. Therefore the starting model (\ref{10})
is equivalent to the sigma model (\ref{12}) only on classical level.
Already at one loop  the auxiliary fields need special care.
At the same time one can use representaition (\ref{12}) to analyze the
general structure of divergences and also to
classify the different versions
of the general model (\ref{9}), (\ref{10}) into several sets.
It is easy to see
that for some versions of (\ref{9}) there is only one significant
big bracket in  (\ref{10}).
For this particular cases we miss one auxiliary field in
(\ref{11}), and thus obtain one less sigma model coordinate in
(\ref{12}), (\ref{13}). The analisys of propagator of the starting
higher derivative model (\ref{2a}) shows that these degenerate models
correspond to \cite{1}
\beq
\Delta = \det
\left(\matrix{
             4 a_1(\phi)   & - a_2(\phi)\cr
             - a_2(\phi)   & a_4(\phi)
\cr}\right)
= (1-uv)^2 =0                                             \label{15}
\eeq
If the condition (\ref{15}) holds and the rank of the matrix in  (\ref{15})
is nonzero then the above scheme has to be modified
because of less amount of the
auxiliary fields, but doesn't fail. The result is similar to that we
have observed in the previous section.
So we see that the higher derivative
terms with $ \left( \nabla\varphi \right)^2$ are of less
importance because they do not give rise to an additional auxiliary dergees of
freedom. It can lead to some difficulty, if we start, for example, with the
theory (\ref{2a}) with $ a_1(\phi) = a_2(\phi) = a_4(\phi) = 0$ when
$a_3(\phi)$ or $a_5(\phi)$ is nonzero. In such theory the target space
metric in the space of scalar field $\phi$, auxiliary field and conformal
factor will be degenerate, and the above scheme does not work. In the
framework of the higher derivative model (\ref{9}) the picture looks as
follows. The inverse propagator of the theory contains only the second
derivative terms, and there are four derivative vertices. As a result the
theory has worst structure of divergences and can be nonrenormalizable
because of possible higher derivative counterterms of $a_1, a_2, a_4$ type.
In terms of the original model (\ref{9})
the one - loop divergences are not related with the functional determinant
of the minimal higher derivative operator (\ref{3f})
but with some complicated nonminimal
second order operator.
So this version of the higher derivative model (\ref{9}) strongly differs
from the others.

\bs

\ni {\bf  Conclusion.} \   We have discussed the
renormalization of a higher derivative dilaton quantum gravity in
two dimensional
space,
and solved some puzzles which have been found in this field.
In particular it was shown that the loop calculations in a harmonic
gauge can be performed in a completely  covariant way. Then we learned
to deal with an auxiliary fields.

It turns out that the general higher derivative model (\ref{9}) can
be reduced to $D = 4$
nonlinear sigma model with the second derivatives only.
In this approach the sigma model coordinats are dilaton scalar field,
conformal
factor of the metric and two auxiliary fields, which correspond to higher
derivatives in the original formulation.
On quantum level the auxiliary fields need some special procedure
to be applied, and then the results of calculations are the same as
in original higher derivative model.

Our consideration may be relevant for the
study of massive modes in string theory and also for the higher derivative
$4d$ quantum gravity.
In this case (as well as in any dimension different from $2$) the curvature
tensor is not defined completely by scalar curvature. In fact it
reflects the existence of spin two states. That is why in four dimensional
theory one can not remove highed derivative fields introducing auxiliary
scalars. However it is quite possible to develop the "second order formalism"
for the fourth derivative gravity, introducing auxiliary tensors of second
rank. If these auxiliary fields are treated in a correct way ( that doesn't
look difficult) then we can obtain new frame for the study of
the general higher derivative $4d$ gravity. Perhaps such
description of the theory will be useful for the better understanding
of the important problem of unitarity.

\vspace{3mm}

\noindent{\bf Acknowledgments}
Author acknowledge the stimulating conversation with S. Odintsov, and
the correspondence with S. Naftulin, who have provided me by useful
explanations of Ref. \cite{naft}.

I am greateful to M. Asorey, J.L. Alonso and to
whole Department of Theoretical Physics at the University of
Zaragoza for warm hospitality, and also to H.S. Song and C. Lee
for  warm hospitality in the Center for Theoretical Physics at Seoul
National University during my visit there.
This work has been
supported  in part by the RFFR (Russia), project
no. 94-02-03234, and by ISF (Soros Foundation), grant RI1000.

\end{document}